\documentclass[prl, showpacs,twocolumn]{revtex4}
\usepackage{amssymb}
\usepackage{graphicx}%
\begin{document}
\title
{Surface pinning in amorphous ZrTiCuNiBe alloy}
\author{V.\,D.\,Fil$^1$, D.\,V.\,Fil$^2$, Yu.\,A.\,Avramenko$^1$,
A.\,L.\,Gaiduk$^1$, and W.\,L.\,Johnson$^3$}
\affiliation{%
$^1$B.\, Verkin Institute for Low Temperature Physics and
Engineering National Academy of Sciences of Ukraine, Lenin av. 47
Kharkov 61103, Ukraine\\ $^2$Institute for Single Crystals,
National Academy of Sciences of Ukraine, Lenin av. 60, Kharkov
61001, Ukraine\\ $^3$California Institute of Technology, Pasadena,
CA 91125}
\begin{abstract}
We have measured the amplitude and the phase of an electromagnetic
(EM) field radiated from superconductor (amorphous ZrTiCuNiBe
alloy) in the mixed state due to interaction of the flux lattice
with an elastic wave. The results undoubtedly point to an
essential contribution of a surface pinning into the flux lattice
dynamics. We propose a model that describes  radiation of EM field
from superconductors with non-uniform pinning.  The model allows
to reconstruct the viscosity and the Labush parameters from the
experimental data. The behavior of the Labush parameter can be
qualitatively explained in terms of the collective pinning theory
with the allowance of thermal fluctuations.
\end{abstract}


\maketitle

Soft type-II superconductors  were studied intensively during last
four decades with the goal to investigate various aspects of
vortex matter dynamics (see the comprehensive review by Brandt
\cite{1}). Nevertheless, many questions that require further
investigation remain. Among these questions is the problem of
relation between the surface and the bulk pinning. Up to now the
dynamics of the vortex state in such an exclusively non-uniform
situation was studied by the surface impedance method \cite{2}. In
this paper on the example of the amorphous
Zr$_{41.2}$Ti$_{13.8}$Cu$_{12.5}$Ni$_{10}$Be$_{22.5}$ alloy we
demonstrate the abilities of a new method based on excitation of
the vortex lattice oscillations by a high-frequency sound wave.

The essence of the method is the following. A superconductor
situated in lower half-space ($z<0$) is subjected by a constant
magnetic field ${\bf H}\ \|\ z$. A transverse elastic wave
propagating along ${\bf H}$ and polarized in $x$ direction
produces transverse (with respect to ${\bf H}$) oscillations of
the vortex lattice caused by pinning forces and viscous friction
forces, and, consequently, induces electromagnetic (EM) field. An
antenna receives EM field (with $E_y$ and $H_x$ components)
radiated through the elastically free surface of the sample (the
surface perpendicular to the direction of propagation of the
elastic wave). While similar experimental setup was already
applied for the study of type-II superconductors \cite{4,5}, the
key new point is measuring both the amplitude and the phase of EM
field (more accurately, the changes of these quantities).

In the uniform case and in the local limit ($q\gg l^{-1}$, $q$ is
the wave number and $l$ is the mean free path) the components of
EM field at $z=0$ are given by a simple expression \cite{6,7}
\begin{equation}\label{1}
  H_x=E_y=\frac{\dot{u}(0)}{c} H \frac{k^2}{q^2+k^2},
\end{equation}
where $u(z)= u_0 \cos qz e^{i \omega t}$ is the elastic
displacement, $k^2$ is the square of complex wave number of EM
field in a conductor, and $c$ is the light velocity.

Eq. (\ref{1}) is applicable for normal as well as for
superconducting state of metal. In normal state $k^2=k_n^2=4\pi i
\omega \sigma_0/c^2$ ($\sigma_0$ is the static conductivity in
normal state) is imaginary valued quantity. In  superconducting
state at small magnetic fields ($H\sim H_{c1}$)
$k^2=k_s^2=\lambda_L^{-2}$ ($\lambda_L$ is the London penetration
length) is real valued. In well-developed Shubnikov state
$k^2=k_m^2=4\pi (i\omega\eta+\alpha_L)/H^2$, where $\eta>0$ is the
viscosity parameter and $\alpha_L>0$ is the Labusch "spring"
parameter. We count the phase $\varphi$ of EM field from its value
at small ($H= H_{c1}+0$) magnetic field. As follows from Eq.
(\ref{1}), the phase is positive at all $H>H_{c1}$ and it
approaches $\varphi_n<90^\circ$ (the phase in normal state) at
$H\to H_{c2}$.

\begin{figure}
\includegraphics[height=6cm, angle=-90]{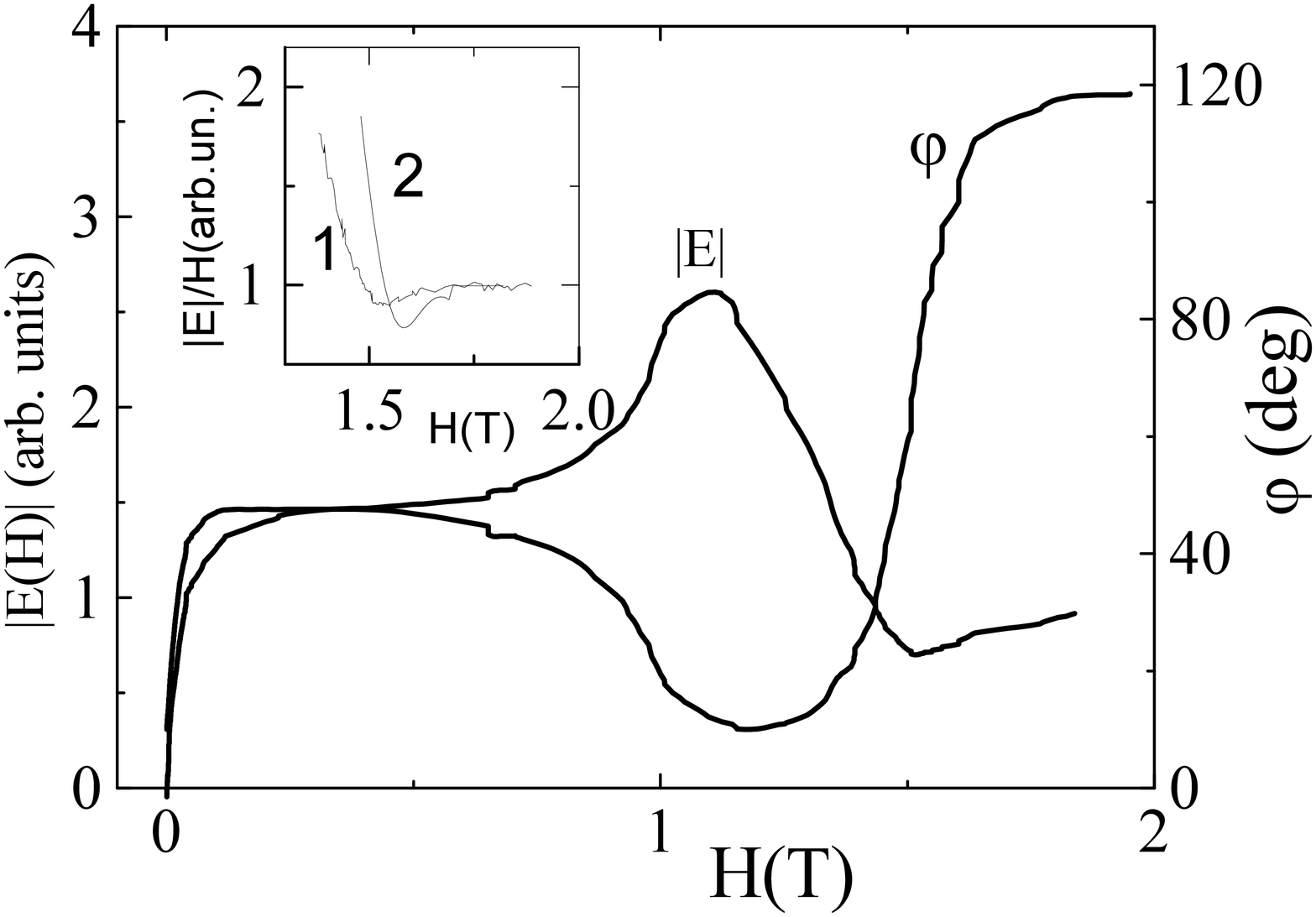}
 \caption{\label{f1} Amplitude and phase of EM field measured for the sample
 with imperfect surface. Inset - behavior of $|E|/H$ near $H_{c2}$
 (1 - experiment, 2 - computed from Eq. (7) with $\alpha_L^{eff}$ (see below))}
\end{figure}

The initial goal of our experiment was to obtain experimentally
the dependencies $\eta(H)$ and $\alpha_L(H)$ using Eq. (\ref{1})
as it was done before for MgB$_2$ \cite{8}. We use the working
frequency $\omega/2\pi\sim 55$ MGz. Details of the measuring
procedure are described in \cite{9}. The results of our first
experiments are presented in Fig. \ref{f1}.

Prior discussing these results let us give the values of
parameters for the material under study that important for further
analysis. The measured d.c. resistivity $\rho\approx 200$
$\mu\Omega\cdot$cm is practically temperature independent. The
Alfer-Rubin effect (quadratic dependence of the sound attenuation
coefficient on the magnetic field \cite{10}) allows to determine
the parameter $\beta=|q^2/k_n^2|=91\pm 1$, in excellent
coincidence with the value of $\rho$ measured (the sound velocity
was found in \cite{11}). The Hall constant, measured at room
temperature, is positive and very small: $R_H=\frac{1}{e n
c}=(3.2\pm 0.4)\cdot 10^{-25}$ CGS units, that yields rather high
density of the carriers $n\approx 2.2 \cdot 10^{23}$cm$^{-3}$.
Taking $m=m_e$ we obtain the relaxation time $\tau \approx
0.8\cdot 10^{-16}$s and for the Fermi velocity $v_F\sim 10^8$cm/s
we find $l\sim 10^{-8}$cm, which is close to the dielectrization
threshold. It was shown before \cite{12} that the alloy under
study belongs to the family of weak coupling superconductors with
the standard BSC energy gap $\Delta(0)\approx 1.75 T_c$
($T_c\approx 0.85$K).  In the dirty limit the formula for the
penetration depth \cite{12a}  can be rewritten as
$\lambda_L^{-2}=k_n^2 \frac{\Delta}{i\omega} \tanh
\frac{\Delta}{2T}$, that yields $\lambda_L=3\cdot 10^{-4}$ cm at
$T=0.4$K. The coherence length calculated from $H_{c2}$ is
$\xi$(0.4K)$=1.4\cdot 10^{-6}$ cm and the Ginzburg-Landau
parameter is $\kappa=\lambda_L/\xi\sim 200$.

As follows from Eq. (\ref{1}), for $\beta$ given above one could
expect $\varphi_n\approx 90^\circ$ that contradicts with the
result presented in Fig. \ref{f1} ($\varphi_n\approx 120^\circ$).
Let us show that the discrepancy found can be accounted for
lowering of conductivity near the surface of the sample caused by
imperfectness of the surface.

We imply the following model dependence of the conductivity on
$z$: $\sigma_0(z)=\sigma_{0v}[1-p\ \exp(z/z_\sigma)]$ (with $p\leq
1$ and $z_\sigma>0$). Then in normal state the electrodynamic
equation for the EM field has the form (see \cite{7}):
\begin{equation}\label{2}
  \frac {d^2 \tilde{E}}{ d \zeta^2}- a(\zeta) \tilde{E}=a(\zeta)
  \cos (\zeta),
\end{equation}
where $a(\zeta)=a[1-p\ \exp(\zeta/\zeta_0)]$, $a=k_{nv}^2/q^2$,
$k_{nv}^2=4\pi i \omega \sigma_{0v}/c^2$, $\zeta_0=q z_\sigma$.
Here we use the dimensionless variables $\tilde{E}=E c/(i\omega
u_0 H)$ and $\zeta=q z$. The boundary conditions for Eq. (2) are
that $|\tilde{E}(\zeta)|$ is finite at all $\zeta$ and $d
\tilde{E}/d\zeta\Big|_{\zeta=0}=0$. The latter condition is due to
continuity of EM field on the conductor-vacuum interface and is
valid with the accuracy $\delta/\lambda_{EM}\lesssim 10^{-4}$
($\delta$ is the skin depth and $\lambda_{EM}$ is the wavelength
of EM field in vacuum).

The solution of Eq. (\ref{2}) can be expressed through the Bessel
functions of the 1-st kind of complex order $\nu=2\zeta_0
\sqrt{a}$ on complex variable $t(\zeta)=2\zeta_0 \sqrt{p
a}\exp(\zeta/2\zeta_0)$. The field at the conductor-vacuum
interface is determined by the expression:
\begin{equation}\label{3}
  \tilde{E}(z=0)= \left(\frac{d J_\nu[t(\zeta)]}{d
  \zeta}\Bigg|_{\zeta=0}\right)^{-1}\int_{-\infty}^{0} d\zeta J_\nu
  [t(\zeta)] a(\zeta) \cos \zeta.
\end{equation}

One can evaluate Eq. (\ref{3}) using the expansion of $J_\nu(t)$
in series in $t$ and integrating each term analytically. For the
case of interest it is enough to take into account first two terms
of the expansion.

Analysis of Eq. (\ref{3}) shows that the phase $\varphi_n$
increases under increasing  $p$. In our case of extremely dirty
conductor ($|a|\sim 10^{-2}$)  and for $\zeta_0\sim 1$ we find
that a rather small lowering of surface conductivity ($\sim 10
\%$) comparing to the bulk one results in increasing of
$\varphi_n$ up to 120$^\circ$.

\begin{figure}
\includegraphics[height=7cm, angle=-90]{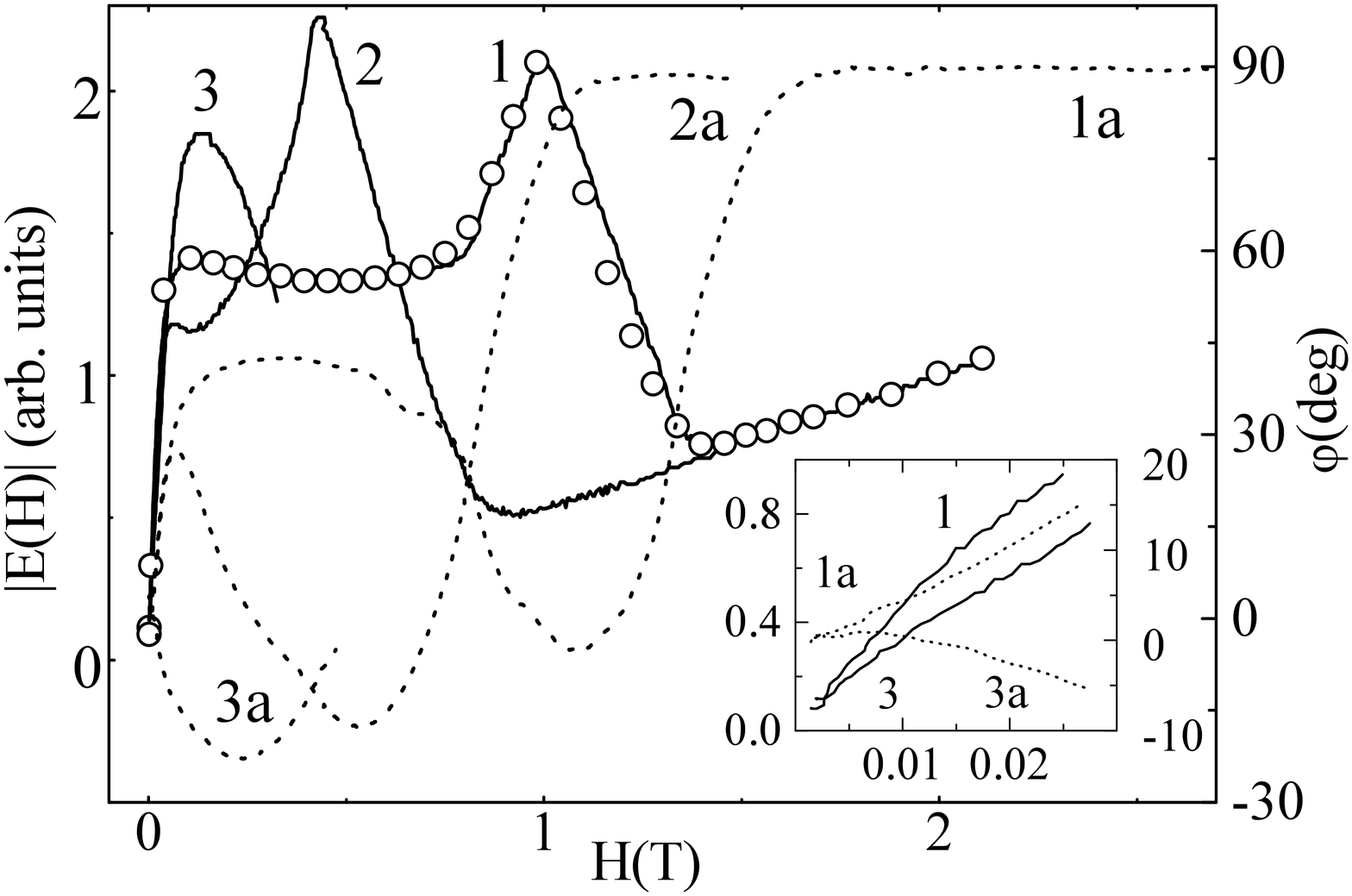}
 \caption{\label{f3} Amplitude (solid curves 1,2,3) and phase (dotted curves 1a,2a,3a)
 of EM field measured for the sample with perfect surface at $T=0.4, 0.69, 0.8$K,
 correspondingly (for $T=0.8$K the results for the normal state are not shown).
 Circles - dependence $|E(H)|$ measured under decreasing of $H$.
 Inset - the same dependencies in the interval of small magnetic fields (all
 notations are the same as in the main plot).
  }
\end{figure}

Removing the layer $\sim 50$ $\mu$m from the surface of the sample
by additional polishing with a fine powder (the size of grains
$\sim 1\div 2$ $\mu$m) we managed to eliminate the lowering of
surface conductivity. The results of measurements (at various $T$)
are shown in Fig. \ref{f3}. One can see that in this case
$\varphi_n\approx 90^\circ$.
The main peculiarity of the dependencies observed is that  at all
temperatures the phase $\varphi$ becomes negative at intermediate
magnetic fields. This effect cannot be described by Eq. (1) under
any reasonable variation of $\eta$ and $\alpha_L$ with $H$.

In what follows we will argue that the effect observed can be
explained by non-uniform pinning near the surface of the sample
that takes place even in a situation with uniform conductivity
(and, consequently, the uniform parameter $\eta$). Since peculiar
behavior of the phase are observed in samples having less perfect
as well as more perfect surface  we consider the non-uniform
pinning  be an intrinsic property of the surface of the material
under study.

EM field in the mixed state is described by the system of
equations, namely, the Maxwell equation, the matter equation, that
determines the value of the current in the two-fluid model, and
the equation of motion for the vortex lattice \cite{7}:
\begin{eqnarray}\label{5}
&&  \frac{d^2 {\bf E}}{d z^2}=\frac{4\pi i \omega}{c^2} {\bf
j}=k_s^2 ({\bf E}+ \frac{1}{c} {\dot {\bf u}_v}\times {\bf
  H}), \\ \label{5c}
 && \frac{1}{c} {\bf j}\times {\bf H} + i\omega\eta ({\bf u}-{\bf
  u}_v)+\alpha_L ({\bf u}-{\bf
  u}_v)=0 ,
\end{eqnarray}
where ${\bf u}_v$ is the displacement of the vortex lattice. Here
we neglect the normal component of the current.

Non-uniform pinning can be modelled by $z$-dependent Labush
parameter. We specify this dependence  as
$\alpha_L(z)=\alpha_{Lv}+ \alpha_{Ls}\exp(z/z_p)$. For the
magnetic field satisfied the inequality $k_s^2 \frac{H^2}{4\pi}\gg
|i\omega\eta+\alpha_L|$ we obtain from (\ref{5}), (\ref{5c}) the
equation for EM field that coincides in form with Eq. (\ref{2})
with the same boundary conditions and the quantity $a(\zeta)$
defined as
\begin{equation}\label{8}
  a(\zeta)=\frac{i\omega\eta+\alpha_{Lv}}{q^2\frac{H^2}{4\pi}}+
  \frac{\alpha_{Ls}}{q^2\frac{H^2}{4\pi}}\exp\frac{\zeta}{\zeta_0}\equiv
  A+C\exp\frac{\zeta}{\zeta_0},
\end{equation}
where $\zeta_0=q z_p$.

Taking into account first two terms in the expansion of the Bessel
function we obtain from Eq. (\ref{3}) the approximate expression
\begin{equation}\label{9}
  \tilde{E}(0)=\frac{\sqrt{A}\frac{A}{1+A}+C\frac{\zeta_0}
  {1+\zeta_0^2}}{\sqrt{A}+C\zeta_0}.
\end{equation}
Eq. (\ref{9}) can be used to reconstruct the dependencies
$\eta(H)$ and $\alpha_L(H)$ from the experimental data presented
in Fig. \ref{f3}, if the parameters $\zeta_0$ and
$\alpha_{Lv}/\alpha_{Ls}$ are specified. These parameters are not
known, but there are  limitations on the choice of them. We
require the functions $\eta(H)$ and $\alpha_L(H)$ be smooth and
real valued. Numerical analysis shows that these requirements
satisfy for $\alpha_{Lv}\ll \alpha_{Ls}$ (it means that the bulk
pinning can be neglected) and $\zeta_0\lesssim 0.5$.

\begin{figure}
\includegraphics[height=6cm, angle=-90]{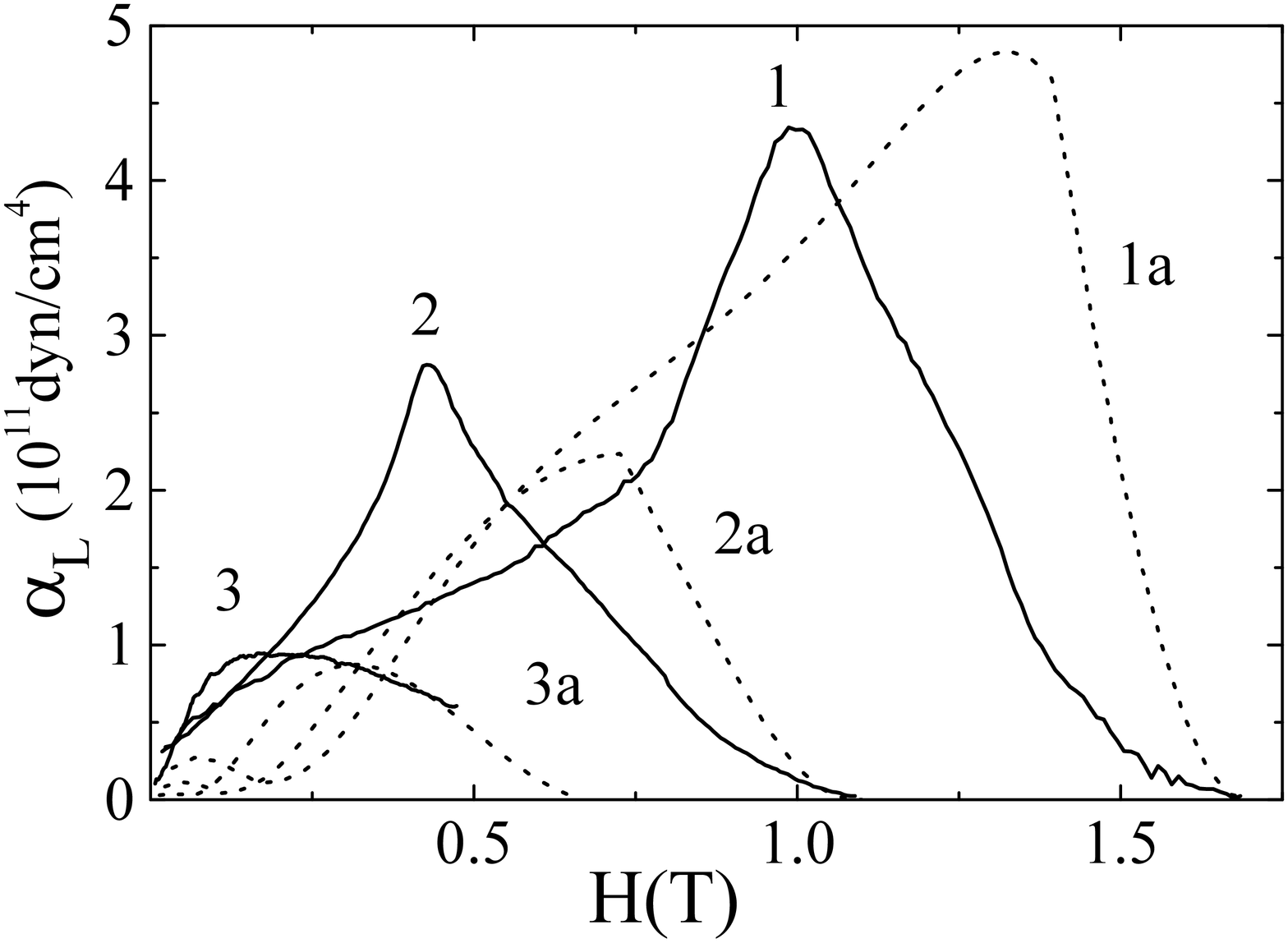}
 \caption{\label{f4} Solid curves 1,2,3 - values of
 $\alpha_L$ reconstructed  from the data in Fig, \ref{f3} with using
 Eq. (\ref{9}); dashed curves 1a, 2a, 3a -
 computed dependence $\alpha_L^{eff}(H)$ (see text)
 at $T=0.4, 0.69, 0.8$K, correspondingly.}
\end{figure}

The dependencies $\alpha_{Ls}(H)$  obtained for $\zeta_0=0.5$ are
shown in Fig. \ref{f4}.

 There are physical
reasons for $\zeta_0$ to be close to $\zeta_0=0.5$. One can see
from Eq. (\ref{9}) that smaller values of $\zeta_0$ correspond to
larger values of $\alpha_{Ls}$. But down to the lowest
temperatures available in our experiment we do not find (see Fig.
\ref{f3}) any signs of freezing of the magnetic flux (the
irreversibility line). It means that the pinning is quite weak and
it is caused, most probably, by point defects.  The estimates (see
further) based on the collective pinning (CP) theory \cite{13}
show that the values of $\alpha_{Ls}$ for $\zeta_0=0.5$ (presented
in Fig. \ref{f4}) are close to maximal possible value of this
parameter.

It is interesting to note that $\zeta_0\sim 0.5$ corresponds to
$z_p\sim \lambda_L$. Probably, it is a fingerprint of that the
image forces are responsible, in some part, for a formation of
$z$-profile of the pinning potential.

\begin{figure}
\begin{center}
\includegraphics[height=6cm, angle=-90]{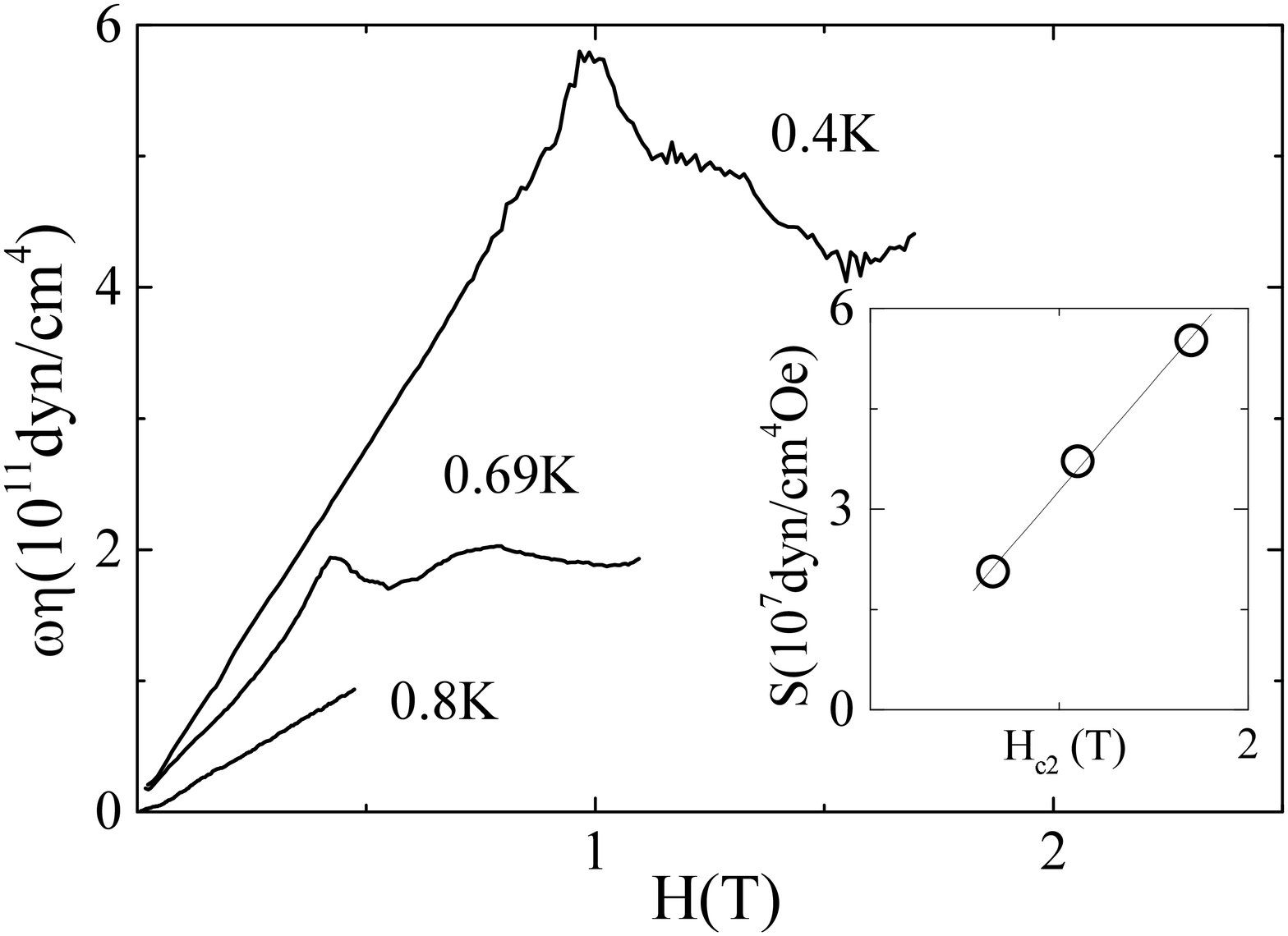}
 \caption{\label{f5} Values of $\omega\eta$ reconstructed  from the data in Fig, \ref{f3} with using
 Eq. (\ref{9}). Inset - the slop $S$ of $\omega\eta$ for intervals of linear increase of $\omega\eta$ with $H$
 (circles); solid line -linear fit.}
\end{center}
\end{figure}

The dependencies $\eta(H)$ are unsensitive to the choice of
$\zeta_0$ and linear in a rather wide interval of $H$   with the
slope $S\propto H_{c2}$ (Fig.\ref{f5}). In this interval they have
the form $\eta(H)=(2\pm 0.05)H H_{c2}\sigma_n/c^2$.

The CP theory \cite{13} allows to compute the field and
temperature dependence of the Labush constant  using only one free
parameter. This parameter is connected with the quenched disorder
of the vortex lattice caused by a random pinning potential. It is
convenient to choose a field $H_{sv}\leq 0.5 H_{c2}$ as such a
parameter\cite{14}. This field separates the regime of single
vortex pinning (SVP) from the regime of bundle vortex pinning
(BVP) . In the SVP regime ($H\leq H_{sv}$ or $H\geq
H_{c2}-H_{sv}$) the value of $\alpha_L$ is given by the expression
$\alpha_L\approx C_{66} a_v^{-2}(H_{sv})$, where
$C_{66}=\frac{\Phi_0 H}{(8\pi
\lambda_L)^2}\left(1-\frac{H}{H_{c2}}\right)^2$ is the shear
modulus of the vortex lattice, $a_v(H)=\sqrt{\Phi_0/H}$ is the
vortex lattice constant, and $\Phi_0$ is the quantum of magnetic
flux. In the BVP regime ($H_{sv}\leq H \leq H_{c2}-H_{sv}$) the
Labush parameter is $\alpha_L\approx C_{66} R_c^{-2}(H)$, where
the collective pinning radius $R_c$ is given by the formula
\cite{mik}
\begin{equation}\label{10}
  R_c(H)\approx a_v(H)\exp \left\{\frac{1}{2}\left[
  \left(\frac{H(H_{c2}-H)}{H_{sv}(H_{c2}-H_{sv})}\right)^{\frac{3}{2}}-1\right]\right\}.
\end{equation}
The expressions presented determine the dependence $\alpha_L(H)$
almost symmetric relative to the $H=0.5 H_{c2}$ line with the
shape varied from the bell shape at $H_{sv}\sim 0.5 H_{c2}$ to the
double-hump shape at $H_{sv}\ll H_{c2}$.

One can see from Fig. \ref{f4} that while the value of
$\alpha_{L}$ obtained from the experimental data  is of the same
order as one given by the CP theory estimates, but the dependence
$\alpha_{L}(H)$  is strongly asymmetric relative to the $H=0.5
H_{c2}$ line. We consider that such  behavior of $\alpha_L$ are
caused by thermal fluctuations.

To evaluate the effect of thermal fluctuations one should add a
random force $f_L$ into Eq. (\ref{5c}) \cite{16}. Strictly
speaking, the returning force $j H/c=(H^2/4\pi) \partial^2
u_v/\partial z^2$ in Eq. (\ref{5c}) has to be found from joint
solution of the system (\ref{5}), (\ref{5c}), but for the
estimates one can replace $\partial^2/\partial z^2$ with $-q^2$.
In this case Eq. (\ref{5c}) can be rewritten in the form of
equation of diffusion of the Brownian particle
\begin{equation}\label{11}
  \tilde{\eta} \frac{\partial w}{\partial t}=\frac{\partial}{\partial w}(\tilde{V}_0
  +\tilde{V}_u+\tilde{V}_p)+f_L,
\end{equation}
where $w=u-u_v$, the "tildes" indicate that the corresponding
quantities are given per one diffusing "particle": $\tilde{y}=y
\Phi_0 L_c/H$ ($L_c$ is the collective pinning length), $V_0= (q^2
H^2/8\pi) w^2$, $V_u=(q^2 H^2/4\pi) u w$, and $V_p$ is the pinning
potential. The latter is modelled by a three-well potential
\begin{equation}\label{12}
  V_p=\frac{\alpha_L }{2}\cases{(w+d)^2& $w\leq -d/2$ \cr
  w^2 & $|w|\leq d/2$ \cr (w-d)^2 & $w\geq d/2$ }
\end{equation}
where $d$ is the distance between minima of the pinning potential.
The linearized Fokker-Planck equation that corresponds to Eq.
(\ref{11}) with the pinning potential (\ref{12}) has the exact
solution in terms of hypergeometric functions. At $\alpha_L/(q^2
\frac{H^2}{4\pi})\ll 1$ the following simple estimate for the
averaged displacement of the vortex lattice is found $\langle u_v
\rangle \approx u \cdot 4 \pi (i\omega\eta+\alpha_L^{eff})/(q^2
B^2)$, where $\alpha_L^{eff}=\alpha_L
 (1-\frac{4}{\sqrt{\pi}}c
   e^{-c^2})$ and $c^2=\frac{q^2 H \Phi_0 L_c}{2\pi T} d^2$.
 One can see
from the comparison of this formula with Eq. (\ref{1})  that the
thermal fluctuations may result in an essential reduction of the
effective Labush parameter remaining the viscosity parameter
unchanged. For general case the dependencies $\alpha_{L}^{eff}(H)$
obtained from the solution of the Fokker-Planck equation are shown
in Fig. \ref{f4}. To achieve semi-quantitative agreement between
the theoretical results and the experimental data at $T=0.4$K we
choose $H_{sv}\approx 0.15 H_{c2}$ and $d\approx 5\cdot
10^{-7}$cm. Other important points of the fitting procedure are
the following: I)To agree the maximum value of $\alpha_L^{eff}$
(at 0.4K) with the experimental one we take for $\lambda_L$ that
enter into equation for $C_{66}$ the value
$\lambda_L(0.4K)=1.2\cdot 10^{-4}$cm. The standard temperature
dependence of $\lambda_L$ \cite{12a} is implied. II) We set $L_c=
a_v(H)$. III) The quantity $d$ is assumed temperature dependent
($d(T)\propto \xi(T)$). IV) We imply that temperature dependence
of $H_{sv}$ is determined by the $\delta T_c$ pinning:
$H_{sv}/H_{c2} \propto (1-(T/T_c)^2)^{-1/3}$ \cite{14}. Such an
assumption looks quite reasonable if one takes into account that
there is a superconducting phase with higher $T_c$ on the surface
of our sample \cite{12}.

One can see from Fig. \ref{f4} that computed temperature and field
dependencies of  $\alpha_L$ are  in qualitative agreement with the
experimental results.

It is interesting to note that if we substitute $\alpha_L^{eff}$
into Eq. (\ref{9}) we obtain $|E|/H$ that has a local minimum near
$H_{c2}$. Such a minimum is observed sometimes in our experiment
(Fig. \ref{f1}) as well as in \cite{4}. Physically this minimum is
connected with the screening of EM field radiated from dipper
regions of the sample by the surface layer with reduced
penetration depth. Reminiscence of this minimum is also present in
Fig. \ref{f3}:  under transition from the normal to mixed state
the increase of $|E|$ begins only after substantial lowering of
$\varphi$.

In conclusion, we propose the new method of investigation of
dynamical characteristics of the vortex matter that consists in
measuring the amplitude and the phase of EM field radiated from
the sample under excitation of the vortex lattice oscillations by
elastic wave. It is established that unusual behavior of the
amplitude and the phase of  EM field are accounted for the surface
pinning. The parameters measured (viscosity coefficient and Labush
parameter) are in good agreement with the estimates obtained from
the theories of vortex matter dynamics.

This study is supported in part by CRDF Foundation (Grant No
UP1-2566-KH-03) and by Ukrainian Government Foundation for Basic
Research (Grant No. 0207/00359). We would like to thank
G.P.Mikitik for helpful discussions.

\end{document}